\documentclass{PHYEAUTH}
\usepackage{graphicx}
\usepackage{amsmath}
\usepackage{amssymb}
\newcommand{\nc}{\newcommand}
\nc{\be}{\begin{equation}}
\nc{\ee}{\end{equation}}
\nc{\bea}{\begin{eqnarray}}
\nc{\eea}{\end{eqnarray}}
\nc{\bean}{\begin{eqnarray*}}
\nc{\eean}{\end{eqnarray*}}
\nc{\mb}{\mbox}
\nc{\rnc}{\renewcommand}

\begin{document}

\begin{frontmatter}

\title{
Numerical Investigation on Asymmetric Bilayer System\\ at Integer Filling Factor
%Instructions for Preparation of EP2DS-15 Manuscript
}

\author[address1]{K. Nomura\thanksref{thank1}},
\author[address1]{D. Yoshioka}
\author[address2,address3]{T. Jungwirth}
and
\author[address2]{A.H. MacDonald}

%\author[address1]{Famous~Professor\thanksref{thank1}} 

\address[address1]{Department of Basic Science, University of Tokyo, Komaba, Tokyo 153-8902, Japan}

\address[address2]{Department of Physics, University of Texas, Austin, Texas 78712 }

\address[address3]{Institute of Physics ASCR, Cukrocarnicka 10,162 53 Praha 6, Czech Republic}
\thanks[thank1]{
Corresponding author. 
E-mail: nomura@toki.c.u-tokyo.ac.jp}

\begin{abstract}
Deformation of the easy-axis ferromagnetic state in asymmetric bilayer systems are investigated numerically. Using the exact diagonalization the easy-axis to easy-plane ferromagnetic transition at total filling factor 3 or 4 is investigated.
At still higher filling, novel stripe state in which stripes are aligned in the vertical direction occurs. The Hartree-Fock energies of relevant ordered states are calculated and compared. 
\end{abstract}

\end{frontmatter}

%[main text]
\section{Introduction}

The quantum Hall systems with internal degree of freedom such as spin or layer index exhibit various fascinating phenomena, symmetry breaking, and topological excitations\cite{eis2}.
In a single-layer system all spins are aligned at integer filling even in the absence of the Zeeman splitting due to the exchange interaction.
Similarly, in a double-layer system without spin degree of freedom, pseudospins are aligned when the total filling factor is an integer. Here the pseudospin $\uparrow$ and $\downarrow$ assign electron in the top layer and bottom layer, respectively.
The case of the layer spacing $d\rightarrow 0$ is equivalent to a single-layer system with spin.
 The ground state of the integer filling ferromagnetic state can be written in the form;
\begin{equation}
      |\Psi[{\bf n}]\rangle= \prod_{m=1}^N c_{m{\bf n}}^{\dag} |0\rangle
\end{equation}
where $ c_{m{\bf n}}^{\dag} =\cos(\theta/2)c_{m_\uparrow}^{\dag}+e^{ i \phi}\sin(\theta/2)c_{m\downarrow}$, $\phi$ is the azimuth and $\theta$ is the declination in the pseudospin space. Here and in the following we neglect the electrons in lower Landau levels if exist. The expectation value of the total pseudospin is given by $\langle {\bf S} \rangle=\langle\Psi[{\bf n}]|\sum_m c_{m\sigma}^{\dag}\frac{\overrightarrow {\sigma}_{\sigma\sigma'}}{2} c_{m\sigma'}|\Psi[{\bf n}]\rangle=(\frac{N}{2}){\bf n}$ where $N=N_{\uparrow}+N_{\downarrow}$ is the total number of electrons.
If the direction  $\bf n$ can be arbitrary in the pseudospin space, which is achieved in the limit of $d\rightarrow 0$ in the double-layer system and single-layer system without the Zeeman splitting, the ground state is the isotropic ferromagnetic state.
On the other hand, at finite layer spacing $d$ in double-layer system, electrostatic charging energy is induced and the pseudospins are confined in the XY-plane in the pseudospin space\cite{moon}. This state is called the XY or easy-plane ferromagnetic state.
Recently, however, the quantum Hall systems with Ising type or easy-axis ferromagnetic anisotropy draw a lot of attentions\cite{ising-mono,ising-bi,rezayi}.
The observed hysteretic transport and resistance spikes indicate Ising-like ferromagnetic state\cite{spike-ex,spike-th1,spike-th2}.
 In this paper we investigate the features of the anisotropic ferromagnetic states and their deformations induced by increasing of the layer spacing $d$.

\begin{figure}[b]
%h=here, t=top, b=bottom, p=separate figure page
\begin{center}\leavevmode
\includegraphics[width=0.9\linewidth]{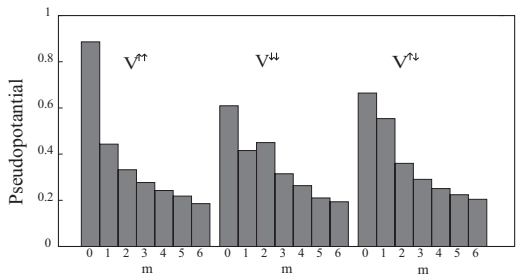}
\caption{Pseudopotentials (a) $V^{\uparrow\uparrow}_m$,(b)
 $V^{\downarrow\downarrow}_m$, and (c) $V^{\uparrow\downarrow}_m$ in the
 case of $N_L^{\uparrow}=0$, $N_L^{\downarrow}=1$. }
\label{figurename}
\end{center}
\end{figure}

\section{Ising quantum Hall ferromagnet}
 The Ising type ferromagnetic states are realized when the two Landau levels of different Landau indices have combined filling factor unity\cite{ising-mono}. 
The anisotropic features are determined by the electron-electron interaction.
The total Hamiltonian is given by
\begin{equation}
H=\sum_{i<j}\sum_{m\sigma\sigma'}V_{m}^{\sigma\sigma'}P_{ij}[m;\sigma\sigma']
\end{equation}
where $ P_{ij}[m;\sigma\sigma']$ projects out components with relative angular momentum $m$ of the electron $i$ and $j$ with pseudospin $\sigma$ and $\sigma'$, and the pseudopotential $V_m^{\sigma\sigma'}$'s are given by 
$V_m^{\sigma\sigma'}=\int_0^{\infty}qdqV_{\sigma\sigma'}(q){\rm L}_m(q^2)e^{-q^2}$ \cite{haldane}. For the isotropic ferromagnetic states $V_m^{\sigma\sigma'}$ are independent of $\sigma$ and $\sigma'$.
On the other hand, when the electrons with $\uparrow$ pseudospin are in the $N_L^{\uparrow}$th Landau level and $\downarrow$ in the $N_L^{\downarrow}$th, $V_m^{\sigma\sigma'}$ significantly depend on $\sigma$ and $\sigma'$.
If $V_{m\ge 1}^{\uparrow\downarrow}$ is large compared with $V_{m\ge
1}^{\uparrow\uparrow}$ and $V_{m\ge 1}^{\downarrow\downarrow}$, the
$\uparrow\downarrow$ channel with relative angular momentum $m$ is not
favored energetically. So the ground state should be $\Pi_m
c_{m\uparrow}^{\dag}|0\rangle$ or $\Pi_m
c_{m\downarrow}^{\dag}|0\rangle$. Actully, as shown in Fig.1, such a
situation realizes at $N_L^{\uparrow}=0$, $N_L^{\downarrow}=1$.

\section{The effect of electrostatic energy}

Next let us consider the effect of increasing the layer separation $d$ from zero.
It might be considered that $V_{m\ge1}^{\uparrow\downarrow}$ decrease as $d$ is increased and the $\uparrow\downarrow$ channels become to be energetically favored.
In a previous work Jungwirth and MacDonald\cite{jungmac} discussed this problem based on the Hartree-Fock theory and found that the transition occurs at $d/l\simeq0.5$ in the case of $N_L^{\uparrow}=0, N_L^{\downarrow}=1$. Now we present an exact diagonalization data of twelve electrons system in Fig.2.
The energies with $n_z=(N_{\uparrow}-N_{\downarrow})/N=0$ and 1 are plotted.
When $d$ is smaller that $d_c\simeq 0.45$ $n_z=\pm1$ state is lowest in energy, while when $d$ exceeds $d_c$ the state with $n_z=0$ becomes lowest.
The $n_z=0$ state is uniform and has a large value of the susceptibility $\chi_{xx}=\frac{d\langle S_x \rangle}{d\Delta_x}$ where $\Delta_x$ is source field along to the $x$-axis. Then this state might be the XY-ferromagnet like state. 
These results are quite consistent with previous Hartree-Fock investigation\cite{jungmac}.

\begin{figure}[h]
%h=here, t=top, b=bottom, p=separate figure page
\begin{center}\leavevmode
\includegraphics[width=0.8\linewidth]{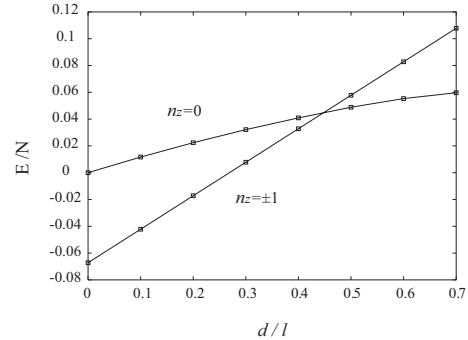}
\caption{The lowest energies of $n_z=0$ states and $n_z=\pm 1$ states as functions of layer spacing $d$.
There is a level crossing between these states. The number of electrons is $N=12$.}\label{figurename}
\end{center}
\end{figure}

\section{Stripe states}
 A domain wall states are stabilized in the small $d/l$ regime in higher levels systems.
Because the electrons cluster on a stripe region in a rectangular system, we call this state as staggered stripe state.
However, some nontrivial aspects are found when the layer spacing still increases to $d/l\ge 0.5$, the situation changes qualitatively\cite{nomura}. The stripe formation occurs not as staggered but as vertical at $N_L^{\uparrow}=0$, $N_L^{\downarrow}=2$ as shown in Fig.3. 

\begin{figure}[t]
%h=here, t=top, b=bottom, p=separate figure page
\begin{center}\leavevmode
\includegraphics[width=0.82\linewidth]{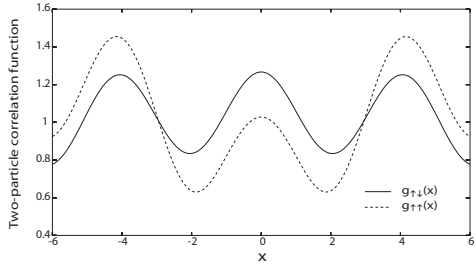}
\caption{Section of the two-particle guiding center correlation function $g_{\uparrow\downarrow}(x)$ (solid line) and $g_{\uparrow\uparrow}(x)$ (dotted line) at the boundary of the rectangular system at $y=L_y/2$.}\label{figurename}
\end{center}
\end{figure}

\section{Mean-field theory for bilayer systems }
%\subsection{Style}

The spatially orderd states such as the stripe state are discribed by the density operator defined as
\begin{equation}
  \hat{\rho}_{\sigma}({\bf q})=\sum_X c_{\sigma X+q_y/2}^{\dag}c_{\sigma X-q_y/2}^{} e^{-iq_xX-q^2/4}.
\end{equation}
Now we inspect the possibility of the stripe states in asymmetric bilayer systems in the framework of the Hartree-Fock approximation\cite{cote,brey,demler}.
To find the HF ground state we have to find the solution with minimum energy.
In what follows we consider a limited number of physically expected solutions and compare their energies.
These solutions are the following:\cite{cote}

\vspace{0.3cm}
\begin{enumerate}
\item  Uniform state: 
$\langle \hat\rho(0) \rangle\neq 0;$ in this state translational symmetry is not broken.
\item  XY ferromagnet:$\langle \hat\rho(0) \rangle\neq 0,\ \langle \hat{S}_+(0) \rangle\neq 0;$ in this state translational symmetry is not broken but pseudospins are polarized in the XY plane in the pseudospin space
\item  Ising ferromagnet: $\langle \hat\sigma(0) \rangle\neq 0$;
 in this state translational invariance is not broken and 
all of the electrons are in only one of the layers.
\item Staggered stripe:
$\langle \hat\sigma(m{\bf Q}_0) \rangle\neq 0$; in this state the region of high electron density distribute alternatively in one direction.
\item Vertical stripe:$\langle \hat\rho(m{\bf Q}_0) \rangle\neq 0$; contrary to the staggered stripe state, the stripes
have same phase in their oscillation in each layer.

\end{enumerate}
\vspace{0.5cm}

\begin{figure}[b]
%h=here, t=top, b=bottom, p=separate figure page
\begin{center}\leavevmode
\includegraphics[width=0.8\linewidth]{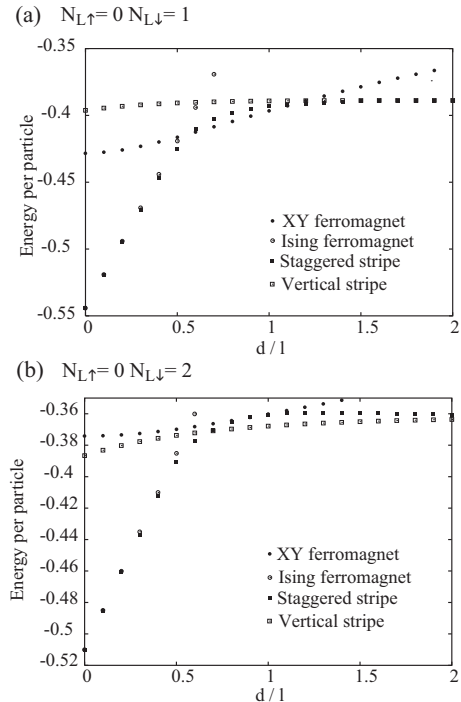}
\caption{Comparison of the Hartree-Fock energy as a function of the layer spacing $d$.}\label{figurename}
\end{center}
\end{figure}

Here we indtoduced the charge density operator and the pseudospin density operator as
$
 \hat{\rho}({\bf q})= \hat{\rho}_{\uparrow}({\bf q})+ \hat{\rho}_{\downarrow}({\bf q})$,
$
 \hat{\sigma}({\bf q})= \hat{\rho}_{\uparrow}({\bf q})- \hat{\rho}_{\downarrow}({\bf q}) 
$ and
$
 \hat{S}^+({\bf q})=\sum_X c_{\uparrow X+q_y/2}^{\dag}c_{\downarrow X-q_y/2}^{} e^{-iq_xX-q^2/4}.
$
As shown in Fig.4 our results indicate the staggered stripe state at $d/l<0.5$ and the XY-ferromagnetic state at $d/l>0.5$ in the case of $N_L^{\uparrow}=0, N_L^{\downarrow}=1$. At still large $d/l$ region, the energies of the staggered and vertical stripe state are quite closed that means the interlayer correlation is quite weak. In the small $d/l$ regime the period of the staggered stripes is much larger than the magnetic length, and at the limit $d/l\rightarrow 0$ the period diverges. Then this phase is understood as a domain state of the Ising ferromagnet. In the case of $N_L^{\uparrow}=0, N_L^{\downarrow}=2$ the vertical stripe state is stabilized instead of the XY ferromagnet. These results are consistent with our numerical diagonalization investigations.

\begin{figure}[t]
%h=here, t=top, b=bottom, p=separate figure page
\begin{center}\leavevmode
\includegraphics[width=0.9\linewidth]{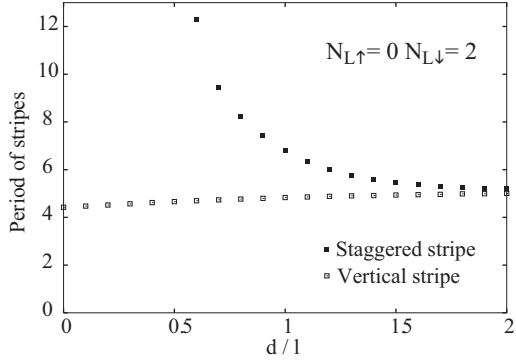}
\caption{Period of the stripes in the unit of the magnetic length $l$ as a function of the layer spacing $d$.}\label{figurename}
\end{center}
\end{figure}

\section{Discussion}

We investigated the many-body ground state in the asymmetric bilayer systems as a function of the layer separation and Landau level index.
We found not only known phases such as the easy-plane ferromagnetic state and the domain wall state but also the vertical stripe in which the electrons distribute in-phase between the layers. The origin of such a peculiar correlation is the difference of the Landau level indices between the layers\cite{nomura}. 
As shown in Fig.6, the interlayer intaraction for the case $N_L^{\uparrow}=0$, $N_L^{\downarrow}=2$ is attractive around
$q=1.5/l$ in which the stripe state is stabilized.
%It is useful to study transport property to see the nature of the ground state experimentally. Recently an observation of anisotropic transport in a single quantum well system was reported\cite{ex2}. The well is not so wide to be regarded as a bilayer system. The exact diagonalization analysis with realistic Coulomb potentials done by Rezayi et al. showed that the ground state is a uniform Ising ferromagnet not a stripe state\cite{rezayi}.
 %It should be interesting to study transport phenomena in a bilayer system.
 We can expect manifestations of such a novel stripe state and
 anisotropic transport phenomena in a ultra high mobility sample.

\begin{figure}[h]
%h=here, t=top, b=bottom, p=separate figure page
\begin{center}\leavevmode
\includegraphics[width=0.9\linewidth]{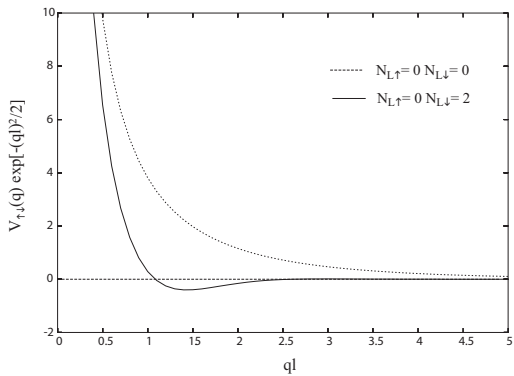}
\caption{Wave number dependence of interlayer interaction
 $V^{\uparrow\downarrow}$ in the case of $N_L^{\uparrow}=0$,
 $N_L^{\downarrow}=0$ and $N_L^{\uparrow}=0$, $N_L^{\downarrow}=2$.}
\label{figurename}
\end{center}
\end{figure}

\end{document}